\documentclass[a4paper,11pt]{article}
\pdfoutput=1 

\usepackage{jheppub} 

\usepackage[T1]{fontenc} 

\newcommand{\be}{\begin{equation}}
\newcommand{\ee}{\end{equation}}
\newcommand{\ba}{\begin{aligned}}
\newcommand{\ea}{\end{aligned}}

\def \sha{{\,\amalg\hskip -3.6pt\amalg\,}}

\begin{document}
\thispagestyle{empty}

\null\vskip-12pt \hfill CERN-PH-TH-2015-183\\
\null\vskip-12pt \hfill LAPTH-043/15 \\

\vskip2.2truecm
\begin{center}
\vskip 0.2truecm {\Large\bf
{\Large Hexagon OPE Resummation and Multi-Regge Kinematics}
}\\
\vskip 1truecm
{\bf J.~M. Drummond${}^{1,2,3}$, G. Papathanasiou${}^{3}$ \\
}

\vskip 0.4truecm
{\it
${}^{1}$ School of Physics \& Astronomy, University of Southampton\\
Highfield, Southampton, SO17 1BJ, United Kingdom\\
\vskip .2truecm                        }
\vskip .2truecm
{\it
${}^{2}$ Theory Division, Physics Department, CERN\\
CH-1211 Geneva 23, Switzerland\\
\vskip .2truecm                        }
\vskip .2truecm
{\it
${}^{3}$ LAPTh, CNRS, Universit\'{e} de Savoie\\
F-74941 Annecy-le-Vieux Cedex, France\\
\vskip .2truecm                        }
\vskip .2truecm
\end{center}

\vskip 1truecm
\centerline{\bf Abstract}

We analyse the OPE contribution of gluon bound states in the double scaling limit of the hexagonal Wilson loop in planar $\mathcal{N}=4$ super Yang-Mills theory. We provide a systematic procedure for perturbatively resumming the contributions from single-particle bound states of gluons and expressing the result order by order in terms of two-variable polylogarithms. We also analyse certain contributions from two-particle gluon bound states and find that, after analytic continuation to the $2\to 4$ Mandelstam region and passing to multi-Regge kinematics (MRK), only the single-particle gluon bound states contribute. From this double-scaled version of MRK we are able to reconstruct the full hexagon remainder function in MRK up to five loops by invoking single-valuedness of the results.

\newpage
\setcounter{page}{1}\setcounter{footnote}{0}

\tableofcontents

\section{Introduction}

Light-like polygonal Wilson loops have been the subject of much study recently, in particular because of their relation to scattering amplitudes in the planar limit of $\mathcal{N}=4$ super Yang-Mills theory \cite{Alday:2007hr,Drummond:2007aua,Brandhuber:2007yx,Drummond:2007cf,Bern:2008ap,Drummond:2008aq,Berkovits:2008ic,Mason:2010yk,CaronHuot:2010ek}. The conformal symmetry of the theory and its associated Ward identity \cite{Drummond:2007au} imply that the Wilson loops with four or five sides are essentially trivial, with no interesting dependence on the configuration of the contour. From six points onwards however there is a conformally invariant function of the loop contour which needs to be determined. Here we will focus on the bosonic hexagonal Wilson loop, which corresponds to the six-particle MHV amplitude under the amplitude/Wilson loop duality. The conformally invariant function of the loop is known as the `remainder function' in this case and is a function of three cross-ratios.

One approach to studying the kinematical dependence of light-like Wilson loops is to make an ansatz based on the analytic behaviour of explicit results obtained for terms in the perturbative expansion such as \cite{DelDuca:2010zg,Goncharov:2010jf,CaronHuot:2011ky}. This approach has yielded results up to four loops at six points \cite{Dixon:2011pw,Dixon:2011nj,Dixon:2013eka,Dixon:2014voa,Dixon:2014iba} and up to three loops at seven points \cite{Drummond:2014ffa}. The ans\"atze are given in terms of iterated integrals over words formed from a specified set, or alphabet, of rational functions (known as letters). The alphabets appearing in the explicit two-loop expressions of \cite{CaronHuot:2011ky} were observed \cite{Golden:2013xva} to be described by the $\mathcal{A}$-coordinates of a class of cluster algebras \cite{1021.16017,1054.17024} associated to the Grassmannians $G(4,n)$. Taking this observation as an assumption, together with basic analytic information on the locations of possible cuts of the final expressions produces a rather restrictive ansatz for the relevant functions. Indeed, at three loops and seven points, these analytic assumptions were essentially used to replace the dynamical information of the theory, producing a unique result for the symbol of the heptagon remainder function \cite{Drummond:2014ffa}.

A second very powerful approach to describing the Wilson loops is based on a type of operator product expansion applied to configurations of null Wilson lines \cite{Alday:2010ku,Gaiotto:2011dt,Basso:2013vsa,Basso:2013aha,Basso:2014koa}. In this approach the Wilson loops are given by an infinite sum over excitations of a light-like flux tube.
This sum amounts to an expansion around a collinear limit, and the excitations may also be thought of as insertions of the fields of the theory (gluons, fermions and scalars) on the Wilson line segments that are becoming collinear. The number of excitations or insertions is equal to the order at which these appear in the near-collinear expansion. The spectrum of excitations was calculated exactly in \cite{Basso:2010in} using techniques based on the integrable structure exhibited in the planar $\mathcal{N}=4$ theory. Apart from the spectrum, a set of overlap functions, called pentagon transitions, are required to calculate the near-collinear expansion. In \cite{Basso:2014nra} an all-loop formula was proposed to describe the subset of pentagon transitions consisting of any number of gluons, including multi-particle bound states.

It is interesting therefore to understand how the near-collinear OPE expansion can be resummed into the type of cluster polylogarithmic functions appearing in the perturbative expansions of the null Wilson loops.
As a first step towards the resummation of the OPE series, we develop a systematic procedure for summing the contributions to the OPE of single-particle bound states of an arbitrary number of gluons in the weak coupling expansion of the MHV 6-point (or hexagon) remainder function. Since these states have a restricted dependence on the helicities of the gluons, this can really be thought of as a contribution to the `double scaling' limit of the Wilson loop, where one of the three cross-ratios is taken to zero. In the perturbative regime the double scaling limit is entirely governed by the gluon insertions, i.e. to determine it entirely at each perturbative order one needs all multi-particle bound states of gluons but not the contributions of fermions and scalars \cite{Basso:2014nra}.

Building on the previous works \cite{Papathanasiou:2013uoa,Papathanasiou:2014yva}, our procedure relies on the technology of nested sums \cite{Moch:2001zr}, and in particular its \texttt{nestedsums} C++ library realisation \cite{Weinzierl:2002hv}. Using these algorithms we are able to show that the OPE expansion for single-particle gluon bound states can always be resummed into two-variable polylogarithmic functions based on a five-letter alphabet. These functions may be described as $A_2$ polylogarithms in the cluster algebra language \cite{Golden:2013xva} or polylogarithms on $\mathcal{M}_{0,5}$ in the language of \cite{FBThesis} or as a subset of the two-dimensional polylogarithms of \cite{Gehrmann:2000zt}. In fact we find that only a particular subset of these functions arises, which is consistent with the functions having restricted branch cuts, and indeed with the idea that they are particular limits of hexagon functions \cite{Dixon:2013eka,Dixon:2014voa} describing the full six-point remainder function.

Perhaps more importantly, from the knowledge of the single-particle bound state contributions to the double scaling limit we can produce all but the power-suppressed terms of the MHV hexagon in multi-Regge kinematics corresponding to high energy gluon scattering in the $2\to 4$ Mandelstam region. The kinematic limit on the cross-ratios is formally the same as the soft limit, in which the remainder function vanishes. To obtain a non-trivial dependence in multi-Regge kinematics, it is necessary to analytically continue the remainder function to the $2\to 4$ Mandelstam region, achieved by going around a singularity where one of the cross-ratios vanishes before taking the limit. In \cite{Hatsuda:2014oza} it was observed that this continuation can be carried out order by order in the OPE expansion. Using this property we provide evidence that after analytically continuing and taking the limit, the single-particle bound states we have summed are the only excitations with non-vanishing contributions on a one-dimensional double scaling slice of the full two-dimensional space parameterising multi-Regge kinematics. In other words we find that, after analytic continuation, the contribution from multi-particle bound states of gluons in the double scaling limit is always suppressed when taking the multi-Regge limit. 

Under the well-justified assumption that the natural function space in these kinematics is given by the class of single-valued polylogarithms (or SVHPLs) \cite{Dixon:2012yy}, we may then reconstruct the remainder function in the full parameter space, from its knowledge on the line obtained from the double-scaling limit. We have carried out this procedure explicitly up to five loops.
We emphasise that in our approach the contribution of the scalars and fermion insertions in the OPE expansion is non-vanishing, even after analytic continuation and taking the Regge limit. Indeed one can already see this in the analysis of \cite{Hatsuda:2014oza}. The matter contributions just vanish on the line we are considering, hence we are actually reconstructing their contribution from the gluon bound states by invoking single-valuedness.

Our logic is very much in line with the analysis of \cite{Basso:2014pla} who also reconstructed the multi-Regge limit of the hexagon amplitudes in the $3\to 3$ Mandelstam region from the OPE contributions of just the single-particle gluon bound states. Indeed, starting with the finite-coupling integral expression for the same OPE contribution that we consider at weak coupling, and performing an additional analytic continuation in Mellin space, the authors of \cite{Basso:2014pla} arrived at finite-coupling expressions for the MHV hexagon in multi-Regge kinematics.  As this second continuation is fundamentally non-perturbative, it is quite interesting that in our approach we can trade it for single-valuedness directly at the perturbative level. 


The paper is structured as follows. We begin in section \ref{DSOPE} by describing the OPE expansion of the hexagon Wilson loop in the double scaling limit. Our focus is on the contribution of the single-particle gluon bound state contributions and their resummation order by order in perturbation theory into two-variable polylogarithms of a particular kind, which we perform in section \ref{subsec_singlegluons}. We then describe in \ref{subsec_hexfn} why the class of functions so obtained is consistent with the idea that the full hexagonal Wilson loop remainder function is expressed in terms of hexagon functions.

In section \ref{MRK} we describe the procedure we use to analytically continue the results of the resummation to the double scaling limit of the $2\to 4$ Mandelstam region and then take the limit to (double scaled) multi-Regge kinematics.
Having obtained a particular limit of the result in multi-Regge kinematics we then describe in section \ref{completion} how we can complete our expressions to restore the full kinematical dependence in multi-Regge kinematics by demanding that the result is single-valued. The analysis of this section is also supplemented by appendix \ref{2particle}, where several two-particle bound states are computed and shown not to contribute to MRK. 

Attached to the arXiv submission for this paper are files containing our results for the resummation of the single-particle gluon bound states up to five loops, particular contributions from two-particle gluon bound states contributing to the double scaling limit and the new results at N${}^3$LLA and N${}^4$LLA for the completion of the five-loop remainder function in multi-Regge kinematics.

\section{The Hexagon Wilson Loop OPE}
\label{DSOPE}

\subsection{Preliminaries}\label{sec_prelim}

An operator product expansion (OPE) for light-like Wilson loops was introduced in \cite{Alday:2010ku} and refined in many papers \cite{Gaiotto:2011dt,Basso:2013vsa,Basso:2013aha,Basso:2014koa,Basso:2014nra,Belitsky:2014sla,Belitsky:2014lta,Basso:2014hfa,Belitsky:2015efa}. It describes the near-collinear regime of a particular ratio of light-like Wilson loops, denoted by $\mathcal{W}$. Here we will focus on the hexagonal Wilson loop where the ratio takes the following form,

\begin{figure}
\centering
\includegraphics[width=0.50\textwidth]{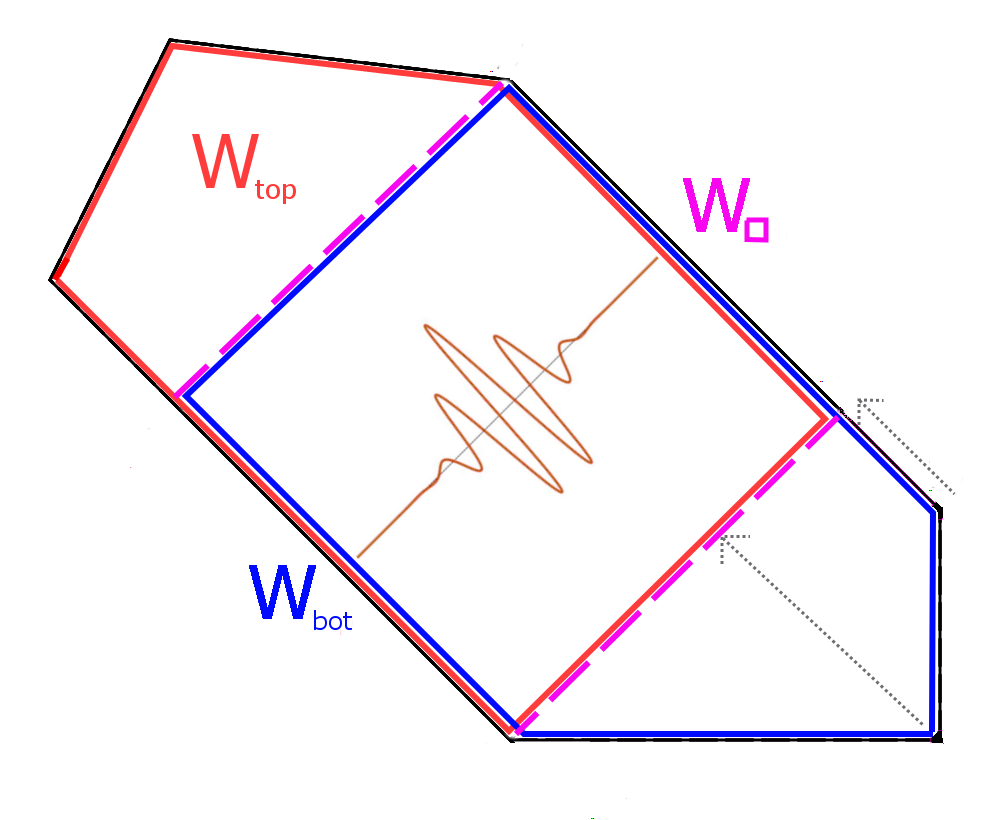}
\caption{Decomposition of the light-like hexagonal Wilson loop in to a top pentagon, bottom pentagon and intermediate square. The collinear limit is indicated by the arrows, and each term in the expansion around it may be mapped to an excitation of an integrable colour-electric flux tube, sourced by the two sides of $W_\Box$ adjacent to the ones becoming collinear.}
\label{fig:hexOPE}
\end{figure}

\be
\mathcal{W} = \frac{ W W_{\Box}}{W_{\rm top} W_{\rm bot}}\,.
\ee
Here $W$ is the hexagonal Wilson loop, while $W_{\rm top}$ and $W_{\rm bot}$ correspond to the pentagonal Wilson loops and $W_\Box$ to the square Wilson loop indicated in Fig. \ref{fig:hexOPE}.

The ratio $\mathcal{W}$ is finite and conformally invariant and is hence a function of the three available conformal cross-ratios,
\be
u_1=\frac{x_{46}^2x_{13}^2}{x_{36}^2 x_{14}^2}\,, \qquad 
u_2=\frac{x_{15}^2 x_{24}^2}{x_{14}^2 x_{25}^2}\,, \qquad
u_3=\frac{x_{26}^2 x_{35}^2}{x_{25}^2x_{36}^2}\,.
\ee

The OPE describes the ratio $\mathcal{W}$ in an expansion around the limit where the two adjacent edges $x_{56}$ and $x_{61}$, of the hexagon become collinear so that $x_{15}^2$ vanishes. The limit is most conveniently parameterised by the variables $\{\tau,\phi,\sigma\}$ which are related to the conformal cross-ratios as follows
\be
\begin{aligned}
u_1& =\frac{1}{2}\frac{e^{2 \sigma+\tau}  \text{sech}\tau}{1+e^{2 \sigma }+ 2\,e^{\sigma -\tau } \cos \phi +e^{-2 \tau }}\,,\\
u_2&= \frac{1}{2} e^{-\tau } \text{sech}\tau  \,,  \\
u_3&= \frac{1}{1+e^{2 \sigma }+ 2\,e^{\sigma -\tau } \cos \phi +e^{-2 \tau }}  \,. \label{crossratios}
\end{aligned}
\ee
The limit $\tau\to\infty$ corresponds to the collinear limit. The quantity $\mathcal{W}$ becomes 1 in the $\tau \rightarrow \infty$ limit, up to exponentially suppressed corrections. The Wilson loop OPE gives a prediction for the form of the exponentially suppressed corrections, namely at $l$ loops
\be\label{Wcol}
\mathcal{W}^{(l)} =\sum_{m=1}^\infty e^{-m\tau}\sum_{p=0}^{[m/2]} \cos [(m-2p)\phi]\sum_{n=0}^{l-1} \tau^n f^{(l)}_{m,p,n}(\sigma)\,.
\ee
Here $[x]$ denotes the integer part of $x$.

The basic idea behind the OPE is to express the bottom part of the Wilson loop as a coherent sum of excitations of the GKP string. These excitations then propagate to the top part of the Wilson loop where they are absorbed. Both the spectrum of excitations \cite{Basso:2010in}, which controls the propagation of states, and overlap functions \cite{Basso:2013vsa}, describing their production and absorption, can be studied at finite 't Hooft coupling using integrability. 

The OPE expansion then has the following schematic form
\be
\label{OPEschematic}
\mathcal{W} = \sum_\psi P(0|\psi) P(\psi | 0) e^{- E_\psi \tau + i p_\psi \sigma + i m_\psi \phi}\,.
\ee
It is a sum over intermediate states $\psi$, weighted by the overlap functions for production and absorption (called `pentagon transitions') $P$ and a factor due the propagation from bottom to top involving the GKP energy (or `twist') $E_\psi$, momentum $p_\psi$ and helicity $m_\psi$. For the purposes of this paper all the relevant quantities are available in the literature and we will describe them in greater detail in the following subsections.


The ratio $\mathcal{W}$ is very simply related to the remainder function $R_6$ via
\be\label{RtoW}
R_6=\log\mathcal{W}-\log\mathcal{W}_{\rm BDS}\,,
\ee
with
\begin{align}
\log\mathcal{W}_\text{BDS}&={\Gamma_\text{cusp}\over4}\{\text{Li}_2\left(u_2\right)-\text{Li}_2\left(1-u_1\right)-\text{Li}_2\left(1-u_3\right)+\log ^2\left(1-u_2\right)\nonumber\\
&\quad-\log \left(u_1\right) \log\left(u_3\right)-\log\left(u_1/ u_3\right)\log\left(1-u_2\right)+\frac{\pi ^2}{6}\}\,,\label{logW_BDS}
\end{align}
and where the overall coefficient $\Gamma_\text{cusp}$ in the latter formula is the cusp anomalous dimension, whose expansion in the coupling
\be\label{eq:g_to_a}
g^2\equiv\frac{\lambda}{(4\pi)^2}=\frac{a}{2}\,
\ee
is as follows,
\be\label{gamma_cusp}
\begin{aligned}
\Gamma_\text{cusp}=\sum_{l=1}^\infty g^{2l}\Gamma_\text{cusp}^l&=4 g^2-\frac{4\pi ^2}{3} g^4 +\frac{44 \pi ^4}{45}g^6-4 \left(\frac{73 \pi ^6}{315}+8 \zeta_3^2\right)g^8\\
&\quad +4 \left(\frac{16 \pi ^2 \zeta_3^2}{3}+160 \zeta_5 \zeta_3+\frac{3548 \pi ^8}{14175}\right)g^{10}+\mathcal{O}\left(g^{12}\right)\,.
\end{aligned}
\ee

In the next section we will describe the `double-scaling' limit of $\mathcal{W}$ (and hence the remainder function $R_6$) which allows us to consider contributions to the sum over states (\ref{OPEschematic}) coming from gluons only.

\subsection{The gluon contributions and the double-scaling limit}

Apart from the collinear limit we described in the previous section, another kinematical limit which will relevant for our discussion is the so-called \cite{Gaiotto:2011dt,Basso:2014nra}
\be\label{doublescaling}
\text{`double scaling limit':}\quad \tau, i \phi\to \infty\quad\,, -\tau+i \phi  \,\,\text{fixed}\,.
\ee
At the level of the OPE expansion (\ref{Wcol}) it amounts to the subset of contributions with $p=0$ and $\cos m\phi\to e^{im\phi}$, which are evidently the only ones surviving the limit. And as far as the cross-ratios are concerned, from (\ref{crossratios}) and (\ref{doublescaling}) we deduce that
\be\label{crossratios_doublescaling}
u_2\to 0\,\qquad u_1\to \tilde u_1=\frac{1}{1+e^{-2 \sigma }+ \,e^{-\sigma -\tau +i \phi} }\qquad u_3\to \tilde u_3=\frac{1}{1+e^{2 \sigma }+ \,e^{\sigma -\tau +i \phi} }\,,
\ee
namely the double scaling limit describes a two-dimensional subspace of the kinematics with vanishing $u_2$, but general values for the remaining cross ratios.

From the point of view of the OPE, this limit is interesting because a restricted set of relatively simpler flux tube excitations contribute to it. In particular, these are the gluon excitations with positive helicity, given that all fundamental scalar, fermion and gluon excitations have helicity (the charge conjugate to $\phi$ in (\ref{OPEschematic})) 0,$\pm 1/2$ and $\pm 1$, all of them have twist 1 at weak coupling, and bound states are formed between gluons with the same helicity, with their charges being just the sum of the charges of their constituents. It is important to note that the twists receive corrections at each order in the coupling while the helicities do not. It is therefore important that, in restricting to the gluon contributions only, we are first expanding perturbatively in the 't Hooft coupling $g^2$ before taking the double scaling limit.
That is, in the limit in question\footnote{Note that although $\mathcal{W}_{BDS}$ has an unphysical branch cut at $u_2=1$, as can be seen from (\ref{logW_BDS}), this is not visible in the double scaling limit, since $u_2\to 0$.}
\be
R_6 \to \widetilde R_6=\log\mathcal{W}_\text{g+}-\log\mathcal{\widetilde W}_{BDS}\,,
\ee
where $\mathcal{\widetilde W}_{BDS}$ is obtained from (\ref{logW_BDS}) after substituting (\ref{crossratios_doublescaling}), and the all-plus gluon contribution is a sum over an arbitrary number of effective particles $N$, each of which may consist of any number of bound gluons  $a_i$, $i=1,\ldots,N$ \cite{Basso:2014nra}\,,
\be
\mathcal{W}_\text{g+}=\sum_{N=0}^\infty \sum_{a_1=1}^\infty\sum_{a_2=1}^{a_1}\dots\sum_{a_N=1}^{a_{N-1}} \prod_{k\neq 0} \frac{1}{n_k!}\, \int  \frac{du_1\dots du_N}{(2\pi)^N}\, \frac{\hat \mu_{a_1}(u_1)\ldots \hat\mu_{a_N}(u_N)}{\prod\limits_{i\neq j} P_{a_i| a_j}(u_i|u_j)} \,.
\label{W6allplusgluons}
\ee 
In the above formula $n_k$ counts the number of bound states made up from $a_i=k$ gluons,
\be\label{mu_hat}
\hat \mu_a(u)\equiv\mu_a(u) e^{-E_a(u) \tau+i p_a(u) \sigma+i a \phi} \,,
\ee
and the energy $E_a$, momentum $p_a$, and measure $\mu_a$ of the $a$-th gluon bound state, as well as its pentagon transition to the $b$-th bound state $P_{a| b}$, are reviewed in appendix \ref{appx_pentagons}.

As a final remark, notice that the weak coupling expansions (\ref{weak_E})-(\ref{weak_P}) of all the different ingredients of (\ref{W6allplusgluons})-(\ref{mu_hat}), imply that the $N$-particle all-plus gluon state starts contributing at order $\mathcal{O}(g^{2N^2})$. Namely one need only consider the states with $N=1$ up to three loops, and $N\le 2$ up to 8 loops. 

\subsection{Resumming all single-particle gluon bound states}\label{subsec_singlegluons}

In this section, we will present a systematic procedure for perturbatively resumming the contribution of all single-particle gluon bound states to the hexagon Wilson loop/scattering amplitude, namely the $N=1$ term in (\ref{W6allplusgluons}), which we may rewrite as
\be\label{W1}
\mathcal{W}_{1}\equiv\sum_{l=1}^\infty g^{2l}\mathcal{W}_{1}^{(l)}\equiv\sum_{a=1}^\infty \int  \frac{du}{2\pi}\, \mu_a(u) e^{-E_a(u) \tau+i p_a(u) \sigma+i a \phi}\,.
\ee
This procedure is an extension of the method developed in \cite{Papathanasiou:2013uoa,Papathanasiou:2014yva} for the evaluation of the individual $a=1$ and $a=2$ terms above. After describing its details, we will apply it in order to obtain explicit for expressions up to $l=5$ loops, and also deduce the relevant class of functions describing this contribution to arbitrary loop order. The reader interested in the final result may jump directly to the discussion around eqs.~(\ref{W1_1loop})-(\ref{W12}) and (\ref{W1_basis}).

We start by expanding the integrand in (\ref{W1}) with respect to $g$, following a well-known procedure that we review towards the end of appendix \ref{appx_pentagons}. This reveals that up to $u$- and $a$-independent factors (e.g. $e^{-\tau}$ or powers of $\tau$ and $\sigma$) which we can factor out, all terms are of the general form
\be
\sum_{a=1} (-e^{-\tau+i \phi})^a\int \frac{du}{2\pi}  \frac{e^{2i u\sigma} \Gamma(\frac{a}{2}+iu)\Gamma(\frac{a}{2}-iu)\prod_{i} \psi^{(m_i)}(1+\frac{a}{2}\pm i u)\prod_{j}\psi^{(m'_j)}(\frac{a}{2}\pm i u)}{\Gamma(a)(u+\frac{i a}{2})^{r} (u-\frac{i a}{2})^{r'}}\,,\label{integrand_weakcoupling}
\ee
for different powers of the denominators $r, r'$ and different products/powers and orders $m_i, m_j'$ of the polygamma functions $\psi$, see eq. (\ref{polygamma}) for their definition.

From (\ref{integrand_weakcoupling}), we can immediately infer that at any loop order the poles of the integrand are located at $u=\pm i(a/2+k)$, $k=0,1,2,\ldots$. After we restrict to $\sigma>0$, we can close the contour of integration on the $u>0$ plane, and trade the integral for a sum over residues by virtue of Cauchy's residue theorem. With the help of the recurrence and reflection relations of the gamma function,
\be\label{Gamma_properties}
\Gamma(z+1)=z \Gamma(z)\,,\qquad\Gamma(1-z) \Gamma(z) = {\pi \over \sin{(\pi z)}}\,,
\ee
as well as analogous equations for polygamma functions arising upon differentiation of (\ref{Gamma_properties}), we arrive to sums over residues which always have the form\footnote{The $k=0$ residue is treated separately, as it is the only case where the denominators have poles, and yields simple sums which may be evaluated exactly as in \cite{Papathanasiou:2013uoa}.}
\be\label{generic_residue_term}
\begin{split}
\sum_{a=1}^\infty (-e^{-\tau +i\phi})^{a}\sum_{k=1}^\infty(-e^{-\sigma})^{a+2k}(-1)^k\frac{\Gamma(a+k)}{\Gamma(a)\Gamma(k+1)}\frac{\prod_{i}\psi^{(n_i)}(a+k)\prod_{j}\psi^{(n'_{j})}(k+1)}{(a+k)^{s} k^{s'}}\,.
\end{split}
\ee
The next step is to reexpress the polygamma functions in terms of $S$- or $Z$-sums \cite{Moch:2001zr} via\footnote{In particular, 
$S(k;m;1)=Z(k;m;1)$ are the generalised harmonic numbers, $\zeta_{m}$ the Riemann zeta function, and $ \gamma_E=-\psi(1)=\simeq 0.577$ the Euler-Mascheroni constant.},
\be\label{psi_to_S}
\begin{aligned}
\psi(k+1)&\equiv\psi^{(0)}(k+1)=-\gamma_E+S(k;1;1)\\
\psi^{(m-1)}(k+1)&=(-1)^{m} (m-1)! [\zeta_{m}-S(k;m;1)]\,,
\end{aligned}
\ee
where 
\begin{align}
S(n;m_1,\ldots,m_j;x_1,\ldots,x_j)&=\sum_{n\geq i_1\geq i_2\geq\ldots\geq i_j\geq1}\frac{x_1^{i_1}}{i_1^{m_1}}\ldots\frac{x_j^{i_j}}{i_j^{m_j}}\,,\label{Ssum}\\
Z(n;m_1,\ldots,m_j;x_1,\ldots,x_j)&=\sum_{n\ge i_1>i_2>\ldots>i_j>0}\frac{x_1^{i_1}}{i_1^{m_1}}\ldots\frac{x_j^{i_j}}{i_j^{m_j}}\,,\label{Zsum}
\end{align}
and replace the products of $S$- or $Z$-sums with the same outer summation index  with linear combinations thereof, simply by nesting the independent summation ranges of each term in the product\footnote{This is nothing but the quasi-shuffle algebra property of these objects.}. For example,
\be\label{qshuffle_example}
\begin{aligned}
\left(\sum_{i_1=1}^n \frac{1}{i_1^{m_1}}\right)\left(\sum_{i_2=1}^n \frac{1}{i_2^{m_2}}\right)&=\left(\sum_{i_1=1}^n\sum_{i_2=1}^{i_1}+\sum_{i_2=1}^n\sum_{i_1=1}^{i_2}-\sum_{i_1=i_2=1}^n\right)\frac{1}{i_1^{m_1}i_2^{m_2}}\\
&=\left(\sum_{i_1=1}^n\sum_{i_2=1}^{i_1-1}+\sum_{i_2=1}^n\sum_{i_1=1}^{i_2-1}+\sum_{i_1=i_2=1}^n\right)\frac{1}{i_1^{m_1}i_2^{m_2}}\,,\\
\end{aligned}
\ee
where the first line yields $S$-sums, and the second line $Z$-sums.

For reasons that will become apparent very shortly, we will choose to replace $\psi^{(m)}(k)$ by $S$- and $\psi^{(m)}(a+k)$ by $Z$-sums respectively. After shifting the summation variable $a\to j=a+k$, partial fractioning in $k$ and expanding, (\ref{generic_residue_term}) splits into terms that look like
\be\label{residuestoSsums}
\begin{split}
\sum_{j=1}^\infty \frac{(-e^{-\tau +i\phi-\sigma})^{j}}{j^{n_1}}Z(j-1;n_2,\ldots;1,\ldots,1)\sum_{k=1}^{j-1}\binom{j-1}{k}\frac{(e^{\tau -i\phi-\sigma})^k}{k^{n'_1}}S(k;n'_2,\ldots;1,\ldots,1)\,,
\end{split}
\ee
where we have combined the gamma functions of (\ref{generic_residue_term}) into a binomial coefficient.

Quite remarkably, the sum in $k$ can be done for any collection of $n_i$, with the help of algorithm C of \cite{Moch:2001zr}. Let us give a simple example where $n'_1=1$ and the $S$-sum is absent, so as to convey the basic idea of the algorithm, which is to appropriately manipulate the expression so that the sum in $k$ can be done by means of the binomial theorem,
\be
\sum_{k=1}^{n}\binom{n}{k}x^k=(1+x)^n\,.
\ee
For the example in question, this entails getting rid of the denominator, roughly speaking by rewriting the summand as the integral of its derivative with respect to $x$,
\be
\sum_{k=1}^{n}\binom{n}{k}\frac{x^k}{k}=\sum_{k=1}^{n}\binom{n}{k}\int_0^x\frac{dx'}{x'}x'^k=\int_0^x\frac{dx'}{x'}\left[(1+x')^n-1\right]\,.
\ee
By changing the integration variable to $y=1+x$, the integrand becomes equal to the sum of the first $n$ terms of a geometric series, so that
\be\nonumber
\sum_{k=1}^{n}\binom{n}{k}\frac{x^k}{k}=\int_1^{1+x}dy\frac{1-y^n}{1-y}=\sum_{k=1}^{n} \left.\frac{y}{k}\right|^{1+x}_1=\sum_{k=1}^{n}\frac{(1+x)^k-1}{k}=S(n;1;1+x)-S(n;1;1)\,,
\ee
where in the last line we used the definition (\ref{Ssum}). By recursively going through the exact same steps, we may similarly obtain
\begin{align}
\sum_{k=1}^{n}\binom{n}{k}\frac{x^k}{k^m}&=\int_0^x\frac{dx_m}{x_m}\int_0^{x_m}\frac{dx_{m-1}}{x_{m-1}}\ldots\int_0^{x_2}\frac{dx_{1}}{x_{1}}x_1^k=\sum_{i_1\ge\ldots\ge i_m\ge 1}^n\frac{1}{i_1}\cdots\frac{1}{i_{m-1}}\frac{(1+x)^{i_m}-1}{i_{m}}\nonumber\\
&=S(n;1,\ldots,1;1\ldots 1,1+x)-S(n;1,\ldots,1;1\ldots 1,1)\,.
\end{align}

In practice, the simplest way do any sum in $k$ as in eq.~(\ref{residuestoSsums}), is to exploit an already existing \texttt{C++} implementation of the algorithm, as part of the \texttt{nestedsums} library \cite{Weinzierl:2002hv} within the \texttt{GiNaC} symbolic computation framework \cite{Bauer20021}\footnote{Alternatively, the \texttt{XSummer} package \cite{Moch2006759} for the \texttt{FORM} symbolic manipulation system \cite{Vermaseren:2000nd} offers exactly the same functionality.}. The relevant command is \texttt{transcendental\_sum\_type\_C}, and we have built an interface that calls it directly from \texttt{Mathematica}, which we use for the remaining manipulations. Combining it also with the \texttt{Ssum\_to\_Zsum} command, the evaluation returns $Z$-sums with outer summation index $j-1$.

Then, after employing quasi-shuffle algebra relations generalising (\ref{qshuffle_example}) in order to eliminate the resulting products of $Z$-sums in favour of their linear combinations, we may immediately evaluate the remaining sum in $j$ in (\ref{residuestoSsums}), in terms of multiple polylogarithms,
\be\label{LitoZ}
\text{Li}_{m_1,\ldots,m_j}(x_1,\ldots,x_j)=\sum_{i=1}^\infty \frac{x_1^{i}}{i^{m_1}}Z(i-1;m_2,\ldots,m_j;x_2,\ldots,x_j)\,,
\ee
see appendix \ref{appx_MPL} for more details on their definition and properties.

The final result is most conveniently expressed in terms of the variables
\be\label{xyvars}
x=-e^{-\tau +i\phi-\sigma}=-\frac{{1-\tilde u_1}-{\tilde u_3}}{{\tilde u_1}}\,,\quad y=1+e^{\tau -i\phi-\sigma}=\frac{{1-\tilde u_1}}{{1-\tilde u_1}-{\tilde u_3}}\,,
\ee
and for example at one loop we have (we denote $H_{m_1,\ldots,m_j}(x)\equiv \text{Li}_{m_1,\ldots,m_j}(x,1,\ldots,1)$ the subclass of harmonic polylogarithms \cite{Remiddi:1999ew}),
\be\label{W1_1loop}
\mathcal{W}_{1}^{(1)}=-\text{Li}_{1,1}(x,y)+2 \sigma  H_1(x)+H_2(x)\,,
\ee
and at two loops,
\begin{align}
\mathcal{W}_{1}^{(2)}=&\tau  \left\{\sigma  \left[-4 \text{Li}_{1,1}(x,y)-4 H_2(x)\right]+4 \text{Li}_{1,1,1}(x,y,1)-4 H_3(x)\right\}+4 \sigma ^2 H_{1,1}(x)\nonumber\\
&+\sigma  \left(-4 H_{1,1,1}(x)-4 \text{Li}_{1,1,1}(x,y,1)+\frac{1}{3} (-2) \pi ^2 H_1(x)-4 H_3(x)\right)\label{W12}\\
&+\frac{1}{3} \pi ^2 H_{1,1}(x)-2 H_{1,3}(x)-2 H_{2,2}(x)-2 H_{3,1}(x)-2 H_{1,1,2}(x)-2 H_{1,2,1}(x)\nonumber\\
&-2 H_{2,1,1}(x)+2 \text{Li}_{1,3}(x,y)+2 \text{Li}_{2,2}(x,y)+2 \text{Li}_{3,1}(x,y)+2 \text{Li}_{1,1,2}(x,1,y)-6 H_4(x)\nonumber\\
&+2 \text{Li}_{1,2,1}(x,1,y)+2 \text{Li}_{2,1,1}(x,1,y)+2 \text{Li}_{1,1,1,1}(x,1,1,y)+4 \text{Li}_{1,1,1,1}(x,y,1,1)\,.\nonumber
\end{align}
As we mentioned at the beginning of this section, we have computed $\mathcal{W}_{1}^{(l)}$ up to $l=5$ loops, and the remaining results may be found in the accompanying file \texttt{allboundstates.m} so as to avoid clutter.

We stress that the procedure we have described can be applied in principle at any loop order, and in fact we can make a precise statement about the particular class of multiple polylogarithms in which $\mathcal{W}_{1}^{(l)}$ lies: From (\ref{residuestoSsums}) it is evident that the leftmost argument of the MPLs (\ref{LitoZ}) will always be $x$. Furthermore, by inspecting the steps of algorithm C in \cite{Moch:2001zr}, we observe that it can only generate $Z$-sums with arguments $y, 1/y$ and 1, such that $y$ and $1/y$ always appear in alternating order with $y$ leftmost, if one removes all arguments equal to 1. Combining these two requirements, we infer that only MPLs of the form
\be\label{Li_basis}
\text{Li}_{m_1,m_2,\ldots}(x,1,\ldots,1,y,1,\ldots,1,1/y,1,\ldots,1,y,\ldots)
\ee
can appear in our results.

The basis of MPLs (\ref{Li_basis}) we have obtained may be expressed more transparently in terms of $G$-functions, whose definitions we review in appendix \ref{appx_MPL}. More specifically, the equivalent $\text{Li}$- and $G$-function representations of MPLs are related by (\ref{LitoG}), and it also proves advantageous to rescale all resulting $G$-functions so that the rightmost argument becomes $x$ as in (\ref{xyvars}), by virtue of the identity (\ref{Grescale}).

In this manner, we finally arrive at the following important conclusion: Apart from explicit factors of $\sigma = -\tfrac{1}{2}\log (x(1-y))$ and $\tau \sim -\tfrac{1}{2} \log u_2$ (in the double scaling limit), the OPE contribution of all single-particle bound states $\mathcal{W}_{1}^{(l)}$ is expressed in terms of 
\be\label{W1_basis}
G(a_1,\ldots,a_n;x)\,,\,\,\, a_i\in\{0,1,1/y\}\, \text{ with }\, a_n\ne 0\,, \text{ and }\,a_j\ne 1/y  \,\text{ if }\, a_1=\ldots =a_{j-1}=0\,, 
\ee
at any loop order $l$. This basis of MPLs has dimension $3^{n-1}$ at weight $n$\footnote{We have a total of $2\cdot 3^{n-1}$ $G$-functions with $a_n\ne 0$, minus $1+\sum_{k=3}^n 2\cdot 3^{k-3}=3^{n-1}$ of them with $a_1=\ldots =a_{j-1}=0$ and $a_j= 1/y$.}, and in fact it turns out to be part of the two-dimensional harmonic polylogarithms (2dHPL) of Gehrmann and Remiddi, which were first introduced in the computation of four-point functions of three on-shell and one off-shell leg at two loops \cite{Gehrmann:2000zt}. 

In the next section, we will explore whether these results are consistent with the fundamental assumption of the hexagon amplitude bootstrap \cite{Dixon:2011pw,Dixon:2011nj,Dixon:2013eka,Dixon:2014voa,Dixon:2014iba} for the class of functions describing six-particle scattering in general kinematics.

Calculating the contribution of multi-particle bound states to the double scaling limit is considerably harder than computing the single-particle bound state contributions. We will leave the resummation over such states to future work. However we can say something about the OPE expansion of certain contributions and this is presented in appendix \ref{2particle}.

\subsection{Hexagon functions in the double-scaling limit}\label{subsec_hexfn}

An important finding of the previous section, is that a particular OPE contribution (\ref{W1}) to the hexagon remainder function $R_6(u_1,u_2,u_3)$ (\ref{RtoW}) is always expressed in the basis (\ref{W1_basis}) at any order in the weak coupling expansion. More specifically, this contribution is the only one that survives in the double scaling limit (\ref{crossratios_doublescaling}) up to 3 loops, and is supplemented by the $N=2$ particle gluonic contributions up to 8 loops, as we noted at the end of section \ref{sec_prelim}. Given that $R_6$ is believed to lie in the space of hexagon functions for general kinematics, here we would like to explain why our results are indeed consistent with the double scaling limit $u_2 \rightarrow 0$ of hexagon functions.

First of all we note that hexagon functions are just a special subset of all iterated integrals based on the nine-letter alphabet $\{u_i, 1-u_i ,y_i\}$ for $i=1,2,3$ described in \cite{Dixon:2011pw}. In other words they are a subset of polylogarithms on $\mathcal{M}_{0,6}$, the moduli space of six points on a Riemann sphere. It is simple to see that in the double scaling  limit $u_2 \rightarrow 0$ these reduce to iterated integrals on the five-letter alphabet\footnote{The fact that hexagon functions reduce to this alphabet in the double scaling limit, and thus approach a subset of 2dHPLs, was noted independently in \cite{ldixonnotes}.} $\{u_1, u_3, 1-u_1, 1-u_3, 1-u_1-u_3\}$, or in terms of the variables $x$ and $y$ defined above, on the alphabet $\{x,y,1-x,1-y,1-xy\}$. Thus the double scaling limit will give polylogarithms on $\mathcal{M}_{0,5}$. A basis for such polylogarithms is given in terms of products of iterated integrals of the form $G(m_y;y)G(m_x;x)$ where the weight vector $m_y$ is made of elements drawn from the set $\{0,1\}$ and the weight vector $m_x$ is drawn from the set $\{0,1,\tfrac{1}{y}\}$. Here we have chosen a specific contour of integration over integrable words in the five letters such that one performs the integration along the $y$-axis first and then in the $x$-direction. We could of course have chosen to do it the other way round, yielding a basis with $x$ and $y$ exchanged.

Hexagon functions, as defined in \cite{Dixon:2013eka}, are a special subset of polylogarithms obeying restrictions on their branch cuts, namely that the locations of the branch points in the Euclidean region occur only at the boundary $u_i=0$ or $u_i=\infty$. At the level of the symbols of hexagon functions this is reflected in the fact that the initial entries are only the $u_i$ and not any of the other nine letters. When taking the double scaling limit, the allowed branch point locations take the form
\be
u_1 \rightarrow \tilde{u}_1 = \frac{1}{1-xy}\,, \qquad u_3 \rightarrow \tilde{u}_3 = \frac{x(1-y)}{1-xy}\,.
\ee
In other words there should be no branch cuts around $y=0$ or $x=1$ at all and the branch cuts around $y=1$ must match those around $x=0$. 

To make the branch cuts around $x=0$ explicit we may use the shuffle relations to extract any trailing zeros from the weight vectors of the $G$-functions with argument $x$. In other words we reexpress the $G$-functions with argument $x$ as linear combinations of $G$-functions which have no trailing zeros in their weight vectors and explicit factors of $\log x$. We may then replace the explicit logarithms of $x$ with logarithms of the product $x(1-y)$ at the cost of redefining the $G$-functions with argument $y$ for each element of the basis. However once we have done so we are no longer allowed any further branch cuts at $y=0$ or at $y=1$. Thus we must have no further factors of the form $G(m_y;y)$ at all.

Thus we find that a basis compatible with the constraints on the branch points at $y=0,1$ is given by all products of the form
\be
\log^p (x(1-y)) G(m_x;x) \, ,
\ee
where there are no trailing zeros in the weight vector $m_x$.

Of course finding a function in the above basis does not necessarily satisfy the remaining constraint, namely that there should be no branch point at $x=1$. Only special linear combinations of functions in the above basis will also satisfy this additional constraint. We have checked that the combinations appearing in the results derived above do indeed obey the property of having no branch points at $x=1$.

Finally let us mention that, as described in (\ref{W1_basis}), we never have a weight vector in the $G$-functions which begins with a string of zeros (of any length) followed by a $\tfrac{1}{y}$. We have verified, by analysing the known four-loop remainder function \cite{Dixon:2014voa}, that this property holds also for the full $R_6^{(4)}$ in the double scaling limit. In other words it holds not only for the single-particle gluon bound state contributions but also for the two-particle ones. This property appears to be an additional constraint on the form of the remainder in the double scaling limit which, at least in the case of the single particle states, we can observe as coming from the structure of the OPE.

\section{From Collinear to Multi-Regge Kinematics}\label{sec_OPEtoMRK}

\subsection{The soft limit, analytic continuation and multi-Regge kinematics}
\label{MRK}

Having obtained expressions for the single-particle gluon bound state contribution to the double scaling kinematics we may now ask about relevant physical limits of our expressions. Since our aim will be to obtain the six-particle scattering amplitude in multi-Regge kinematics we will briefly review the relevant  kinematic limits. We recall that the soft limit and the multi-Regge limit are formally identical in that we send one cross-ratio ($u_3$ say) to one and the other two ($u_1$ and $u_2$) to zero so that the ratios
\be
\frac{u_1}{1-u_3}=\frac{1}{(1+w)(1+w^*)},\qquad
\frac{u_2}{1-u_3}=\frac{w w^*}{(1+w)(1+w^*)}\,
\label{MRLvars}
\ee
are fixed. In the formulas above we have introduced the variables $w$ and $w^*$ which parametrise the possible remaining dependence in the limit. In the soft limit we expect the remainder function to simply vanish. However, after we analytically continue the amplitude to the $2\to 4$ Mandelstam region (obtained by continuing around the singularity at $u_3=0$ via $u_3 \rightarrow u_3 e^{-2 \pi i}$), the above limit is non-trivial and we expect a series of divergent logarithms whose coefficients are functions of the remaining variables $w$ and $w^*$,
\be
R_6^{(l)} \longrightarrow (2 \pi i)\sum_{r=0}^{l-1} \log^r(1-u_3) [g_r^{(l)}(w,w^*) + (2 \pi i) h^{(l)}_r(w,w^*)]\,.
\label{Reggelimit}
\ee
The limit (\ref{Reggelimit}) has been studied in many papers (see for example \cite{Bartels:2008ce,Bartels:2008sc,Lipatov:2010qg,Lipatov:2010ad,Bartels:2010tx,Fadin:2011we,Prygarin:2011gd,Lipatov:2012gk, Bartels:2011ge,Bartels:2013jna,Bartels:2014jya}).

In our preceding discussion from section \ref{DSOPE} we were considering the double-scaling limit where one of the cross-ratios, $u_2$ was taken to zero. This means that starting from the double scaling limit we actually only have access to multi-Regge kinematics in the regime where $w^* \rightarrow 0$ with $w$ fixed (or, switching the helicities of the gluon bound states, to the regime $w\rightarrow 0$ with $w^*$ fixed). We refer to this regime as {\sl double-scaled multi-Regge kinematics}.

We may relate the variables $w$ and $w^*$ to the variables $\sigma$, $\tau$ and $\phi$ via,
\be
w=r e^{i\phi}\,,\qquad w^*=r e^{-i\phi}\,, \qquad r = e^{-\tau-\sigma}\,.
\label{wwstar}
\ee
Note also that the variable $x$ introduced to parametrise the double scaling limit is related to $w$ via $x=-w$ and that therefore $w^* = - x e^{-2i\phi}$ and hence terms power suppressed in the double scaling limit correspond to terms power suppressed in $w^*$.

Now let us turn to the properties of the expressions we have obtained in the various physical limits discussed above.
An obvious property, manifest from the form of the OPE expansion is that all $\mathcal{W}^{(l)}$ vanish in the strict collinear limit $\tau\rightarrow \infty$. Likewise, the expressions we have obtained vanish in the soft limit $x \rightarrow 0$ with $y$ fixed. All of the $G$-functions of argument $x$ are power suppressed as $x \rightarrow 0$ and there is one such $G$ in each term of our results.

The other soft limit $y \rightarrow -\infty$ with $x$ fixed is related by the symmetry $u_1 \leftrightarrow u_3$ to the first. Our expressions also vanish in this limit. To see this however requires taking the limit $y \rightarrow -\infty$ on each of the $G$-functions appearing in our expressions. It is instructive to do this explicitly as we will need to perform exactly the same operation when taking the Regge limit.

In order to carry out the limit $y\rightarrow - \infty$ on our expressions it is first useful to express them in such a way that no $G$-functions are divergent in the limit. To arrange this we use the shuffle relations to unshuffle any zeros or $\tfrac{1}{y}$ appearing at the end of the weight vector so that each $G$-function either has a weight vector ending in a one or there are no ones on the weight vector at all. In the latter case we may rescale the argument by $y$ so that these $G$-functions have weight vectors with entries 0 and 1 and argument $xy$ (i.e. they are HPLs of argument $xy$). For example we write
\be
G(1,0,\tfrac{1}{y};x) = -G(1;xy)G(0,1;x)+G(1;x)G(0,1;xy)+G(\tfrac{1}{y},0,1;x)\,.
\ee
The limit $y\rightarrow-\infty$ may now be taken straightforwardly using e.g. the HPL mathematica package \cite{Maitre:2005uu} to give the limits of the $G$-functions with arguments $xy$ as $y \rightarrow - \infty$. In this way we find that all our resummed single-particle bound state contributions to the double scaling limit vanish in the soft limit $y \rightarrow -\infty$.

The very same limit is required to analyse the multi-Regge limit of the double scaling limit of $\mathcal{W}$. However to obtain a non-vanishing contribution we must first analytically continue to the Mandelstam region by passing round the branch but starting at $u_3 =0$. This is easily achieved since all of the cuts at $u_3=0$ are manifest in the form of our result since they appear as explicit logarithms of $x(1-y) = u_3/u_1$. We simply need to continue these so that $\log x(1-y) \rightarrow \log x(1-y) - 2\pi i$. In so doing we obtain the result for $\mathcal{W}$ in the Mandelstam region in the double scaling limit.

Having analytically continued our results we may then go to multi-Regge kinematics by taking exactly the same soft limit as above, namely $y \rightarrow - \infty$ with $x$ fixed. This time of course we obtain a non-vanishing result, dependent on $x = -w$. Finally, we express our results in a way that will be convenient for us to complete our expression in double scaled multi-Regge kinematics to the full multi-Regge kinematics. In order to do this we note that in the double scaling limit we have
\be
\tau = -\frac{1}{2} \log u_2 = -\frac{1}{2} \log(1-u_3) - \frac{1}{2} \log (w w^*) + \frac{1}{2} \log (1+w) \,
\ee
and 
\be
\sigma = - \tau - \frac{1}{2} \log (w w^*) = \frac{1}{2} \log(1-u_3) - \frac{1}{2} \log(1+w)\,.
\ee
We may use these relations to rewrite all contributions in terms of the variables $w,w^*$ and $(1-u_3)$ which parametrise multi-Regge kinematics. In fact we also rewrite $\log (ww^*) = \log (-w) + \log(-w^*)$ and neglect the divergent logarithm $\log(-w^*)$. This is because these divergent logarithms are tied to the finite parts by the completion to single-valued polylogarithms that we describe in the next section.

The above discussion has focussed on the contribution of the single-particle gluon bound states to double-scaled multi-Regge kinematics. We have also analysed certain contributions from two-particle bound states, as described in appendix \ref{2particle}. We find that after analytic continuation and taking the limit these contributions are all power suppressed. Hence, based on this preliminary analysis it seems that the single-particle states give the only non-vanishing contribution to double-scaled multi-Regge kinematics.

\subsection{Completion to full multi-Regge kinematics from single-valuedness}
\label{completion}

Here we would like to explain why deriving the result in multi-Regge kinematics from the double scaling limit is sufficient to reconstruct the full result in multi-Regge kinematics. We know that in the multi-Regge limit of the remainder function in the $2\to 4$ Mandelstam region we will obtain an expansion in powers of the divergent logarithm which we make take to be $\log(1-u_3)$\footnote{Note that in this section we have changed the expansion parameter $g^2\to a=2g^2$ to reflect most commonly adopted conventions in the subject.},
\be
R_6^{(l)} \longrightarrow (2 \pi i)\sum_{r=0}^{l-1} \log^r(1-u_3) [g_r^{(l)}(w,w^*) + (2 \pi i) h^{(l)}_r(w,w^*)]\,.
\ee
The coefficients of the divergent logarithms are separated into an imaginary part $g^{(l)}_r$ and a real part $h^{(l)}_r$. For each $l$ and $r$ both $g^{(l)}_r$ and $h^{(l)}_r$ are single-valued polylogarithms (or SVHPLs) \cite{Dixon:2012yy}.

The expressions we have obtained from the analytic continuation of the double-scaling limit allow us to obtain the limit of each $g^{(l)}_r$ and $h^{(l)}_r$ as $w^* \rightarrow 0$ with $w$ held fixed\footnote{We may treat $w$ and $w^*$ as independent complex variables to perform this limit.}. We also discard any divergent logarithms of the form $\log w^*$ as we take this limit, keeping only the finite term. From this data we may reconstruct the full dependence of all contributions to the real and imaginary parts in multi-Regge kinematics by invoking single-valuedness of the result.

In order to explain this point we briefly recall the construction of single-valued polylogarithms given in \cite{BrownSV}. We begin with Knizhnik-Zamolodchikov equation in a single complex variable $z$,
\be
\frac{d}{dz} L(z) =  \biggl(\frac{e_0}{z}  + \frac{e_1}{1-z}\biggr) L(z)\,.
\label{KZ1}
\ee
Here $e_0$ and $e_1$ are two free non-commuting generators. We take the solution $L_0$ of (\ref{KZ1}) normalised so that as $z\rightarrow 0$ we have $L_0(z) \sim L_0^{\rm an}(z) z^{e_0}$ where $L_0^{\rm an}(z)$ is analytic around $z=0$ and $L_0^{\rm an}(0)=1$. The solution can be represented as a sum over all words in $e_0$ and $e_1$ with coefficients which are harmonic polylogarithms (i.e. regularised iterated integrals over $d\log z$ and $d\log(1-z)$ away from $z=0$),
\be
L_0(z) = \sum_m m H_m(z).
\ee
Here the sum is over all words $m$ in $e_0$ and $e_1$.

A second solution of the same equation may be taken to be 
\be
L_1(z) = L_0(1-z)|_{e_0\rightarrow -e_1, \,\, e_1 \rightarrow - e_0}\,.
\ee
It is normalised so that $L_1(z) \sim L_1^{\rm an}(z) (1-z)^{-e_1}$ as $z\rightarrow 1$ with $L_1^{\rm an}(z)$ analytic at $z=1$ and $L_1^{\rm an}(1) =1$.

The two solutions are related by parallel transport via a constant series $\Phi$ (the Drinfeld associator),
\be
L_0(z) = L_1(z) \Phi\,.
\ee
Here $\Phi$ is a series in $e_0$ and $e_1$ with coefficients which are (shuffle-regularised) multiple zeta values,
\be
\Phi = \sum_m m \zeta_{\sha}(m)\,.
\ee
Explicitly expanded as a series in $e_0$ and $e_1$ we have
\be
\Phi = 1 + [e_0,e_1]\zeta_2 + ([e_0,[e_0,e_1]]-[e_1,[e_0,e_1])\zeta_3 + \ldots
\ee

We may now consider the analytic continuation of the solution around the singular points at $z=0$ and $z=1$. Under an analytic continuation around $z=0$ (via $z \rightarrow ze^{2\pi i}$),
the monodromy of the solution $L_0$ is explicit, given the asymptotics as $z\rightarrow 0$,
\be
\mathcal{M}_0 L_0(z) = L_0(z) e^{2 \pi i e_0}\,.
\ee
The monodromy of $L_0$ around $z=1$ may be obtained by transporting to $z=1$, where the monodromy of $L_1$ is explicit, and then back again,
\be
\mathcal{M}_1 L_0(z) = L_0(z) \Phi^{-1} e^{-2 \pi i e_1} \Phi\,.
\ee
Now, to construct single-valued polylogarithms in a complex variable $z$, we consider a second `primed' alphabet $e_0'$ and $e_1'$ and form the following series in all four letters $e_0,e_1,e_0'e_1'$,
\be
\mathcal{L}(z,\bar{z}) = L_0(z) \tilde{L}_0'(\bar{z})
\ee
The second factor is built on the primed alphabet while the symbol `$\sim$' means that the words in $e_0'$ and $e_1'$ are reversed with respect to the ones in $e_0$ and $e_1$ appearing in $L_0(z)$. 

Now the series $\mathcal{L}(z,\bar{z})$ will be single-valued if it has no monodromy around $z=0$ or $z=1$. The monodromy around $z=0$ is given by
\be
\mathcal{M}_0 \mathcal{L}(z,\bar{z}) = L_0(z) e^{2 \pi i e_0} e^{-2 \pi i e_0'} \tilde{L}_0'(\bar{z})\,.
\ee
The series will be unchanged if 
\be
e_0' = e_0\,.
\label{e'1}
\ee

Similarly the monodromy around $z=1$ is given by
\be
\mathcal{M}_1 \mathcal{L}(z,\bar{z}) = L_0(z) \Phi^{-1} e^{-2 \pi i e_1} \Phi \tilde{\Phi}' e^{2 \pi i e_1'} (\tilde{\Phi}')^{-1} \tilde{L}_0'(\bar{z})\,,
\ee
where $\Phi'$ is built on the letters $e_0'$ and $e_1'$.
The series is unchanged if
\be
\Phi^{-1} e^{-2 \pi i e_1} \Phi \tilde{\Phi}' e^{2 \pi i e_1'} (\tilde{\Phi}')^{-1}  = 1\,,
\label{e'2}
\ee
which, together with (\ref{e'1}), defines the $e_0', e_1'$ alphabet in terms of $e_0$ and $e_1$. The individual coefficients of words in $e_0$ and $e_1$ in the series $\mathcal{L}(z,\bar{z})$ are then necessarily single-valued polylogarithms $\mathcal{L}_m(z,\bar{z})$,
\be
\mathcal{L}(z,\bar{z}) = \sum_m m \mathcal{L}_m(z,\bar{z}).
\ee
In other words, at a given weight, there are exactly as many single-valued poylogarithms in $z$ and $\bar{z}$ as there are harmonic polylogarithms in a single variable $z$ built on $d \log z$ and $d \log(1-z)$.

Given the asymptotics of $L_0$ defined above, the limit of each $\mathcal{L}_m(z,\bar{z})$ as $\bar{z}\rightarrow 0$ (now treating $z$ and $\bar{z}$ as independent complex variables), is a series in divergent logarithms $\log \bar{z}$ with the coefficient of the finite term being simply $H_m(z)$. In other words there is a unique completion of a given harmonic polylogarithm $H_m(z)$ to a single-valued polylogarithm $\mathcal{L}_m(z,\bar{z})$ such that the finite term in the limit $\bar{z}\rightarrow 0$ is $H_m(z)$. This fact was already noticed and found to be very useful in \cite{Drummond:2013nda} in reconstructing the full form of the three-loop `Easy' integral appearing in the correlation function of four stress-tensor multiplets in $\mathcal{N}=4$ super Yang-Mills theory.

\begin{figure}
\centering
\includegraphics[width=0.70\textwidth]{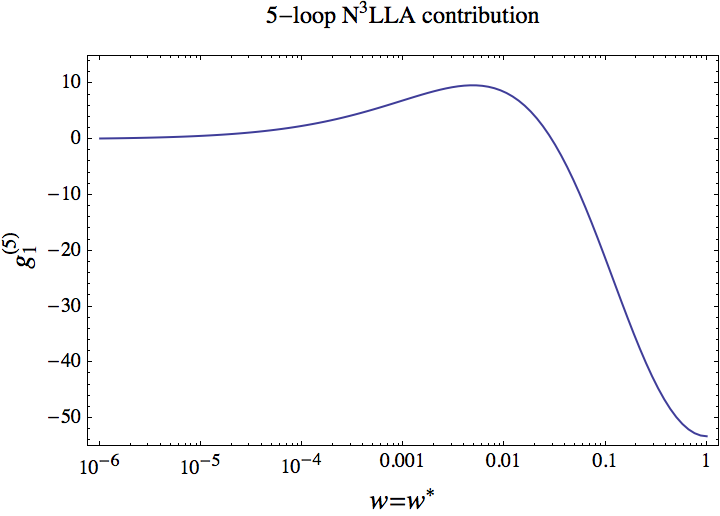}
\caption{Log-linear plot of the imaginary part of the 5-loop hexagon remainder function in multi-Regge kinematics at $(\text{Next-to})^3$-Leading Logarithmic approximation, $g^{(5)}_1$, on the line $w=w^\star$.}
\label{fig:g51plot}
\end{figure}

\begin{figure}
\centering
\includegraphics[width=0.70\textwidth]{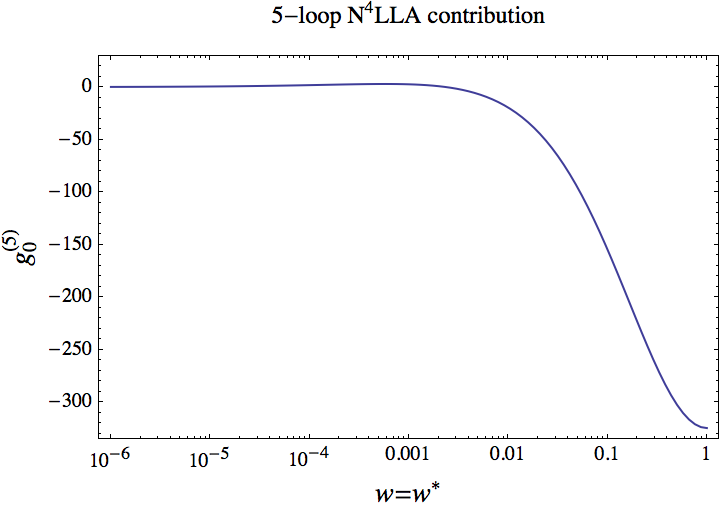}
\caption{Imaginary part $g^{(5)}_0$ of the 5-loop hexagon remainder function in multi-Regge kinematics at $(\text{Next-to})^4$-Leading Logarithmic approximation on the line $w=w^\star$.}
\label{fig:g50plot}
\end{figure}

Thus from the above discussion it is clear that to complete our expressions given in terms of HPLs with argument $(-w)$ into single-valued expressions with the correct limit, we simply replace each instance of $H$ with $\mathcal{L}$. In this way we obtain the full expressions for $g^{(l)}_r$ and $h^{(l)}_r$ dependent on $w$ and $w^*$. We have verified that our results up to five loops reproduce the known expressions derived in \cite{Dixon:2012yy,Dixon:2014voa}. In addition we have obtained the remaining five-loop terms $g^{(5)}_1$ and $g^{(5)}_0$. We also obtain $h^{(5)}_0$ but the real parts may be simply related to the imaginary ones so this does not constitute independent data. We give plots of the new data $g^{(5)}_1$ and $g^{(5)}_0$ along the diagonal line $w=w^*$ in Fig. \ref{fig:g51plot} and Fig. \ref{fig:g50plot}.

\subsection{Comparison with BFKL and the approach of Basso, Caron-Huot and Sever}

From the reconstructed expressions derived in the previous section one can compare to the BFKL formula for the amplitude in multi-Regge kinematics,
\be
\begin{aligned}
e^{R_{6}+i\pi \delta}=\cos \pi \omega_{ab}
+\frac{ia}{2} \sum_{n=-\infty}^\infty (-1)^n \left( \frac{w}{w^*}\right)^{\frac{n}{2}}
\int_{-\infty}^\infty \frac{d\nu}{\nu^2+n^2/4} |w|^{2i \nu} \Phi_{\rm reg} (\nu,n)\\
\times \exp\left[-\omega(\nu,n)\left(\pi i+\log(1-u_3)+\frac{1}{2}\log \frac{|w|^2}{|1+w|^4} \right) \right]\,.
\end{aligned}
\label{BFKLeqn}
\ee
The formula expresses the amplitude in multi-Regge kinematics as a Fourier-Mellin transform of a factorised expression involving the BFKL eigenvalue $\omega(\nu,n)$ and impact factor $\Phi_{\rm reg}(\nu,n)$ which encode all the kinematical dependence of the amplitude. The other quantities in the above formula are given by
\be
\omega_{ab} = \frac{1}{8} \Gamma_{\rm cusp} \log (w w^*)\,, \qquad \delta = \frac{1}{8} \Gamma_{\rm cusp} \log \frac{w w^*}{(1+w)^2(1+w^*)^2}\,.
\ee

In principle, from our reconstructed expressions, we can then find $\Phi_{\rm reg}(\nu,n)$ and $\omega(\nu,n)$ so that they are consistent with the perturbative expansion of the amplitude in multi-Regge kinematics that we reconstructed from the double-scaling limit after analytic continuation. However, since an all-order form for these quantities was obtained in \cite{Basso:2014pla} following a different (though similar) logic, we can simply compare our expressions for the amplitude in multi-Regge kinematics with those obtained from their formula. Doing so we find perfect agreement up to five loops \footnote{Note that in the weak coupling expansions derived in \cite{Basso:2014pla}, one needs to take care that $\nu_{\rm BCS} = 2\nu$.}.

Notice that the reconstruction argument of this section essentially amounts to the claim that the full BFKL expression can be reconstructed just from the knowing the residue at the first pole on the positive $\nu$ axis, assuming that the perturbative expression is given in terms of single-valued polylogarithms in $w$ and $w^*$. Certainly, under this assumption, the simplest way to evaluate any given expression perturbatively is simply to calculate the integral above around the first pole and resum the result as a single-variable function of $w$. Then one may complete this boundary information to a single-valued polylogarithm as described above.

It is very interesting that both here and in \cite{Basso:2014pla} the crucial ingredients were the single-particle bound states of gluons appearing in the OPE expansion. We stress that the other states {\sl do} contribute in multi-Regge kinematics. It is just that we are able to ignore the contributions from the fermions and scalars from the beginning by going to the double scaling limit and then, by extension of our observations, assume that the multi-particle gluon bound states drop out when going to double-scaled multi-Regge kinematics. The contributions of the missing states can then all be reconstructed by appealing to single-valuedness. Note that the issue of identifying the $n=0$ term in the sum (\ref{BFKLeqn}) does not arise in our approach; it follows along with all the others when completing the double-scaled MRK expressions to the full MRK ones.  If we were able to fully justify the fact that the multi-particle gluon bound states drop out from double-scaled MRK we would have proven that the full expression can be reconstructed from the single-particle gluon bound states alone. As the authors of \cite{Basso:2014pla} stress, very similar assumptions have to be made also in their approach. It would be very interesting indeed if some combination of the two arguments could be used to fully prove the BFKL formula for the hexagon in MRK.

\acknowledgments
We'd like to thank B.~Basso, S.~Caron-Huot, L.~Dixon, A.~Sever, E.~Sokatchev and S.~Weinzierl for stimulating discussions, and especially B.~Basso, S.~Caron-Huot and L.~Dixon for comments on the manuscript. The authors are grateful to the Mainz institute for Theoretical Physics (MITP) for the hospitality and support during the course of this work. GP thanks the organisers of the Flux Tubes workshop at Perimeter Institute for the hospitality and support. The work of GP was supported by the French National Agency for Research (ANR) under contract StrongInt (BLANC-SIMI-4-2011).

\appendix

\section{Dispersion relation, measure and pentagon transitions for gluons}\label{appx_pentagons}
Here we review the building blocks of the Wilson loop OPE, coming from all gluonic excitations of the corresponding flux tube \cite{Basso:2014nra}, in slightly adjusted notation. Although these formulas hold at finite coupling, they are particularly suited for the weak coupling expansion, as we will comment below, and display to leading order at the end of this section.

The energy and the momentum of the $a$-th gluon bound state, $a\in \mathbb{Z}^*$ with $a>0$ corresponding to positive and $a<0$ to negative helicity gluons, are given by
\be\label{eq:dispersionreln}
E_a(u)=|a|+4g\, \left[\mathbb{Q} \cdot  \mathbb{M} \cdot  \kappa(a,u)\right]_1\,,\quad\quad p_a(u)=2u-4g\, \left[\mathbb{Q} \cdot  \mathbb{M} \cdot  \tilde \kappa(a,u)\right]_1 \,,
\ee
where $\mathbb{Q}$ is a matrix with elements $\mathbb{Q}_{ij}=\delta_{ij}(-1)^{i+1}i$, $\mathbb{M}$ is related to another matrix $K$,
\be\label{M_and_K}
\mathbb{M}\equiv (1+K)^{-1}=\sum_{n=0}^\infty (-K)^n\,,\quad \quad K_{ij}=2j(-1)^{j(i+1)} \int\limits_{0}^\infty \frac{dt}{t} \frac{J_i(2gt)J_j(2gt)}{e^t-1} \,,
\ee
$J_{i}$ is the $i$-th Bessel function of the first kind, and $\kappa, \tilde \kappa$ are vectors with elements
\be\label{kappa_vectors}
\begin{aligned}
\kappa_{j}(a,u) &\equiv  \int\limits_{0}^\infty \frac{dt}{t} \frac{J_j(2gt)(J_0(2gt)-\cos(ut)\left[e^{t/2}\right]^{(-1)^{j}-|a|+1})}{e^t-1} \\
\tilde\kappa_{j}(a,u) &\equiv  \int\limits_{0}^\infty \frac{dt}{t} (-1)^{j+1}\frac{J_j(2gt) \sin(u t) \left[e^{t/2}\right]^{(-1)^{(j+1)}-|a|+1}}{e^t-1}\,.
\end{aligned}
\ee
The subscripts in (\ref{eq:dispersionreln}) denote that we are only taking the first component of the vector inside the brackets. Moving on, the measure of the $a$-th gluon bound state, and its pentagon transition to a $b$-th bound state are given by 
\begin{align}
\mu_{a}(u)&=F_{a}(u) \,e^{f_3^{(a,a)}(u,u)-f_4^{(a,a)}(u,u)} \,,\label{mu_a}\\
P_{a|b}(u|v)&=F_{a,b}(u,v) \,e^{i f_2^{(a,b)}(u,v)-i f_1^{(a,b)}(u,v)+f_4^{(a,b)}(u,v)-f_3^{(a,b)}(u,v)} \,, \label{P_ab}
\end{align}
where the $f_i^{(a,b)}$ functions are expressed in terms of the previous quantities (\ref{M_and_K})-(\ref{kappa_vectors}),
\be\label{f_ab_functions}
\begin{aligned}
f_1^{(a,b)}(u,v) &= 2 \,\tilde \kappa(a,u) \cdot \mathbb{Q} \cdot  \mathbb{M} \cdot \kappa(b,v)\, ,\quad \,&  f_2^{(a,b)}(u,v) &= 2 \, \kappa(a,u) \cdot \mathbb{Q} \cdot  \mathbb{M} \cdot \tilde\kappa(b,v) \,,\\
f_3^{(a,b)}(u,v)&=2 \,\tilde \kappa(a,u) \cdot \mathbb{Q} \cdot  \mathbb{M} \cdot \tilde \kappa(b,v)\, , &  f_4^{(a,b)}(u,v)&=2 \, \kappa(a,u) \cdot \mathbb{Q} \cdot  \mathbb{M} \cdot \kappa(b,v) \,,
\end{aligned}
\ee
and in addition
\be
F_a(u)=\frac{g^2(-1)^a \Gamma(1+\frac{a}{2}+iu)\Gamma(1+\frac{a}{2}-iu){e^{\int_0^\infty \frac{dt (J_0(2gt)-1)}{t(e^t-1)} (2e^{-at/2}\cos(u t)-J_0(2g t)-1)}}}{\Gamma(a)(x^{[+a]}x^{[-a]}-g^2)\sqrt{((x^{[+a]})^2-g^2)((x^{[-a]})^2-g^2)}} \,  ,
\ee

\begin{align}
&F_{a,b}(u,v)=\sqrt{(x^{[+a]}y^{[-b]}-g^2)(x^{[-a]}y^{[+b]}-g^2)(x^{[+a]}y^{[+b]}-g^2)(x^{[-a]}y^{[-b]}-g^2)} \, \times\\
&\quad\times\, \frac{(-1)^b \Gamma(\frac{a-b}{2}+iu-iv) \Gamma(\frac{a+b}{2}-iu+iv){e^{\int_0^\infty \frac{dt (J_0(2gt)-1)}{t(e^t-1)} (J_0(2gt)+1-e^{-at/2-iu t}-e^{-bt/2+iv t})}}}{g^2 \Gamma(1+\frac{a}{2}+iu) \Gamma(1+\frac{b}{2}-iv) \Gamma(1+\frac{a-b}{2}-iu+iv)} \,,\nonumber
\end{align}
\be
\ba
F_{a,-b}(u,v)&=\frac{{e^{\int_0^\infty \frac{dt (J_0(2gt)-1)}{t(e^t-1)} (J_0(2g t)+1-e^{-at/2-iu t}-e^{-bt/2+iv t})}}}{\sqrt{(1-\frac{g^2}{x^{[+a]} y^{[-b]}})(1-\frac{g^2}{x^{[-a]} y^{[+b]}})(1-\frac{g^2}{x^{[+a]} y^{[+b]}})(1-\frac{g^2}{x^{[-a]} y^{[-b]}})}} \times \\
&  \times\,   \frac{\Gamma(1+\frac{a+b}{2}+iu-iv)}{\Gamma(1+\frac{a}{2}+iu) \Gamma(1+\frac{b}{2}-iv)} \, ,
\ea
\ee
where for the last three formulas we have resticted $a, b>0$, and under the same restriction the remaining cases may be obtained by $F_{-a}(u)=F_{a}(u)$, $F_{-a,-b}(u,v)=F_{a,b}(u,v)$ and $F_{-a,b}(u,v)=F_{a,-b}(u,v)$. Finally, 
\be
x^{[\pm a]}=x(u\pm ia/2)\,,\quad y^{[\pm b]}=x(v\pm ib/2)\,, \quad x(u) = \frac{1}{2}\big(u+\sqrt{u^2-(2g)^2}\big).
\ee

Although the vectors (\ref{kappa_vectors}) and matrices (\ref{kappa_vectors}) are infinite-dimensional, and hence the matrix products (\ref{eq:dispersionreln}) and (\ref{f_ab_functions}) should really be thought of as infinite-dimensional sums at finite coupling, it is possible to show that at weak coupling $K_{ij}$ starts at order $\mathcal{O}(g^{i+j})$. Thus if we wish to obtain the weak coupling expansion up to order $\mathcal{O}(g^{2l})$, we may simply truncate all summations up to their first $i,j=1,2\ldots,2l-1$ terms.

In order to see this, and also perform the expansion in practice, one starts with the Taylor series of the Bessel functions,
\be\label{BesselJ_expansion}
J_i(z)=\sum _{n=0}^{\infty } \frac{(-1)^n }{k! \Gamma (i+n+1)}\left(\frac{z}{2}\right)^{i+2 n}\,,
\ee
and then computes integrals in $t$ by means of
\be\label{psi_integral}
\int_0 ^\infty dt\left(\frac{t^m e^{-zt}}{1-e^{-t}}-\frac{e^{-t}}{t}\delta_{m,0}\right)=(-1)^{m+1}\psi^{(m)}(z)\,,\quad m\ge0\,,
\ee
where $\psi^{(m)}(z)$ is the polygamma function,
\be\label{polygamma}
\psi^{(m)}(z)=\frac{d^{m+1}}{dz^{m+1}}\log \Gamma(z)\,.
\ee
For example, the leading order expressions for all necessary ingredients of the Wilson loop OPE that we obtain in this manner, are for $a,b>0$
\begin{align}
E_a(u)&=a+2g^2[\psi(1+\tfrac{a}{2}+iu)+\psi(1+\tfrac{a}{2}-iu)-2\psi(1)] + \mathcal{O}(g^4)\,,\label{weak_E}\\
p_a(u)&=2u+2ig^2 (\psi(\tfrac{a}{2}+iu)-\psi(\tfrac{a}{2}-iu)) + \mathcal{O}(g^4) \,, \\
\mu_a(u)&=(-1)^a g^2\frac{\Gamma(\frac{a}{2}+iu)\Gamma(\frac{a}{2}-iu)}{\Gamma (a)(u+\frac{ia}{2})(u-\frac{ia}{2})}+\mathcal{O}(g^4) \,,\\
P_{a|b}(u|v)&=\frac{(-1)^b (\tfrac{a}{2}-iu) (\tfrac{b}{2}+iv) \Gamma(\frac{a-b}{2}+iu-iv) \Gamma(\frac{a+b}{2}-iu+iv)}{g^2 \Gamma(\frac{a}{2}+iu) \Gamma(\frac{b}{2}-iv) \Gamma(1+\frac{a-b}{2}-iu+iv)} +\mathcal{O}(g^0) \,.\label{weak_P}
\end{align}
More details about the weak coupling expansion may be found in appendix E of \cite{Basso:2013aha}, as well as appendix A.2 of \cite{Basso:2014nra}).

\section{Remarks on the two-particle gluon bound state contributions}
\label{2particle}
Although we leave the resummation of all two-particle gluon bound states as an exciting open question for future work, in this appendix we shall employ the technology we have developed so far in order to compute the contribution of several such states in the collinear limit (part \ref{appx_2particle1}). The main motivation here is to examine whether these states survive when we transition from collinear to multi-Regge kinematics, and complement the analysis of section \ref{sec_OPEtoMRK} (part \ref{appx_2particle2}) . We gather evidence up to 5 loops that this is not the case, thus allowing us to reconstruct the hexagon remainder function in the latter kinematics from the single-particle gluon bound states, resummed in subsection \ref{subsec_singlegluons}.

\subsection{Evaluation of the integrals}\label{appx_2particle1}
The two-particle gluon bound state OPE contribution corresponds to keeping only the $N=2$ term in eq.~(\ref{W6allplusgluons}), which we may rewrite as
\be
\mathcal{W}_2= \sum_{a=1}^\infty\sum_{b=1}^a \frac{\mathcal{W}_{2[a,b]}}{1+\delta_{ab}}\,,\quad \mathcal{W}_{2[a,b]}\equiv\int_{-\infty}^{+\infty}\int_{-\infty}^{+\infty}  \frac{du dv}{(2\pi)^2}\, \frac{\hat \mu_{a}(u)\hat\mu_{b}(v)}{P_{a|b}(u|v)P_{b|a}(v|u)}\,,
\label{all2gluons}
\ee 
with $a,b$ the number of gluons contained in each of the two bound states (also equal to their twist and helicity), $\delta_{ab}$ a Kronecker delta function, and all ingredients of the OPE integral on the right-hand side contained in appendix \ref{appx_pentagons}. With the help of (\ref{weak_E})-(\ref{weak_P}), at 4 loops the integral becomes
\be
\begin{aligned}
\mathcal{W}^{(4)}_{2[a,b]}&=\int  \frac{du dv}{(2\pi)^2}\frac{\Gamma \left(\frac{a}{2}-i u\right)^2 \Gamma \left(\frac{a}{2}+i u\right)^2 \Gamma \left(\frac{b}{2}-i v\right)^2 \Gamma \left(\frac{b}{2}+i v\right)^2 }{\Gamma (a) \Gamma (b) \Gamma \left(\frac{a}{2}+\frac{b}{2}+i u-i v\right) \Gamma \left(\frac{a}{2}+\frac{b}{2}-i u+i v\right)}\times\\
&\quad\times\frac{(-1)^{a-b}\left[\frac{a-b}{2}-i (u-v)\right] \left[\frac{a-b}{2}+i (u-v)\right]}{(u-\frac{i a}{2})^2 (u+\frac{i a}{2})^2 (v-\frac{i b}{2})^2 (v+\frac{i b}{2})^2}\, e^{-(a+b) (\tau+i\phi)  +2 i \sigma  (u+v)}\,,
\end{aligned}
\ee
where we have replaced four of the gamma functions coming from (\ref{weak_P}) with the help of the identity
\be
\frac{\Gamma \left(1+\frac{a-b}{2}-i u+i v\right)\Gamma \left(1+\frac{b-a}{2}-i v+i u\right)}{\Gamma \left(\frac{a-b}{2}+i u-i v\right) \Gamma \left(\frac{b-a}{2}+i v-i u\right)}={(-1)^{a-b} \left[\tfrac{a-b}{2}-i (u-v)\right] \left[\tfrac{a-b}{2}+i (u-v)\right]}\,,
\ee
which may be derived from (\ref{Gamma_properties}). For specific values of $a, b$, and depending on whether $a/2, b/2$ and $(a+b)/2$ are integer or half-integer, we may similarly replace the remaining gamma functions by virtue of
\be\label{Gamma_products}
\begin{aligned}
\Gamma(n+x)\Gamma(n-x)&=\prod_{m=1}^{n-1}(m+x)(m-x)\frac{\pi x}{\sin(\pi x)}\,,\\
\Gamma(\tfrac{1}{2}+n+x)\Gamma(\tfrac{1}{2}+n-x)&=\prod_{m=0}^{n-1}(\tfrac{1}{2}+m+x)(\tfrac{1}{2}+m-x)\frac{1}{\cos(\pi x)}\,,
\end{aligned}
\ee
for $n$ integer. Finally, we may factorise all $u-$ and $v-$dependence coming from the trigonometric/hyperbolic functions, e.g.
\be
\sinh (\pi u-\pi v)=\sinh (\pi  u) \cosh (\pi  v)-\cosh (\pi  u) \sinh (\pi  v)\,.
\ee

From these considerations, it is clear that the only inseparable dependence on $u-v$ will come from the products in (\ref{Gamma_products}) appearing in the denominator of (\ref{all2gluons}). Given that at higher loops the integrand (\ref{all2gluons}) is dressed by sums of products of polygamma functions with arguments $1+\tfrac{a}{2}\pm u, \tfrac{a}{2}\pm u, 1+\tfrac{b}{2}\pm v$ and $\tfrac{b}{2}\pm v$, this statement holds true independently of the loop order. 

As was first observed in \cite{Papathanasiou:2014yva}, for the case $a=b=1$ these $(u-v)$-dependent terms in the denominator cancel out, and therefore $\mathcal{W}^{(l)}_{2[1,1]}$ always reduces to a sum of two factorised 1-fold integrals, which may be evaluated along the lines of \cite{Papathanasiou:2013uoa}. Here, we find that exactly the same phenomenon occurs also for $a=2$ and $b=1$. 

Although for $a+b\ge 4$ we start dealing with genuine 2-fold integrals, these can still be done by summing over residues, at least at 4 loops. The main subtlety is that we need to treat separately not only the poles at $u=ia/2$ and $v=ib/2$, but also the ones where 
\be
u-v=i\left(-\tfrac{a+b}{2}+1\right),i\left(-\tfrac{a+b}{2}+2\right),\ldots,i\left(\tfrac{a+b}{2}-1\right)\,,\quad u-v\ne \pm i\,\tfrac{a-b}{2}\,.
\ee
These special poles will lead to simple sums, whereas the remaining ones give rise to double sums, which may be brought to a form similar to the one we encountered in \ref{subsec_singlegluons}. Sparing the reader the rest of the details, all in all we arrive at expressions for $\mathcal{W}^{(l)}_{2[1,1]}$ and $\mathcal{W}^{(l)}_{2[2,1]}$ up to $l=5$ loops, as well $\mathcal{W}^{(4)}_{2[3,1]}$ and $\mathcal{W}^{(4)}_{2[2,2]}$, which are included in the accompanying file \texttt{boundstatesN2.m}.

Before moving on to examine the contribution of these states in MRK, let us end this section by noting a peculiar property of the two twist-4 contributions: Each one separately gives rise to polylogarithms with symbol letters $S, 1+S^2$ but also $1-S^2$, namely HPLs with both positive and negative weights. At first this may seem to contradict the expectation discussed in section \ref{subsec_hexfn}, that perturbatively $R_6$, and hence also $\mathcal{W}$, are described by a particular 9-letter alphabet, since the latter only reduces to the letters $S$ and $1+S^2$ in the collinear limit. Quite remarkably however, in their sum $\mathcal{W}^{(4)}_{2[3,1]}+\mathcal{W}^{(4)}_{2[2,2]}/2$, also taking into account the symmetry factor in \ref{all2gluons}, all terms containing the letter $1-S^2$ cancel out, and consistency is restored. This perhaps suggests that it is only the combination of states with the same charges and particle number $N$ that has a physical significance in the OPE.

\subsection{Contribution to MRK}\label{appx_2particle2}
In section \ref{MRK}, we reviewed the analytic continuation which is necessary for obtaining a nontrivial result in MRK, and specialised its form in a regime where it overlaps with the double scaling limit, where our resummed single-particle gluon contribution lives. For the individual two-particle gluon bound states that we computed in the previous part of this appendix, we will need a similar specialisation to a region overlapping with the collinear limit. This was first considered in \cite{Bartels:2011xy}, based on the initial Wilson loop OPE approach \cite{Alday:2010ku}. Following the refinement of the latter approach, this procedure has been revisited and extended in \cite{Hatsuda:2014oza}, which we now briefly review.

In the latter reference, the usual analytic continuation in the cross-ratios was translated to $\tau,\sigma, \phi$ variables (\ref{crossratios}) parameterising the near-collinear limit expansion. In more detail, if we denote $S=e^\sigma$ and $T=e^{-\tau}$, a particular path
\be
\mathcal{C}: (e^{i\chi}S, T, e^{i \chi}\cos \phi+i S^{-1}(T+T^{-1})\sin \chi),\quad \chi \in [0,\pi],
\label{eq:analytic_cont}
\ee
connecting the initial point $(S, T, \cos \phi)$ in the Euclidean sheet for $\chi=0$, to the same point $(-S, T, -\cos \phi)$ in the $2\to 4$ Mandelstam sheet for $\chi=\pi$.

In \cite{Hatsuda:2014oza} it was also observed that at two loops, the analytic continuation for $S>1$ and the collinear limit expansion commute. In other words, first analytically continuing $R_6^{(l)}$ and then expanding the near collinear limit yields the same result as first expanding, and then analytically continuing term by term. This behaviour was then conjectured to hold to all loops, and in fact it can be proven for all polylogarithms on $\mathcal{M}_{0,6}$. In other words if one assumes that $R_6^{(l)}$ is described by hexagon functions, then the commuting of the two procedures follows.

Taking the aforementioned property as granted, it is then easy to obtain the collinear limit expansion of the analytically continued $R_6^{(l)}$. For all (sums of) positive helicity gluon states with given particle number and twist $a$ that we have considered  (as discussed in the last paragraph of section \ref{appx_2particle1}), we find that their kinematical dependence is always a linear combination of terms of the form
\be\label{collexp_genform}
(\log T)^m \left(\frac{e^{i\phi}T}{S}\right)^a S^{2k} (\log S)^j H_{m_1,\ldots,m_r}\left(-1/S^2\right)\,,\quad k=0,1\ldots a\,,\quad m_i>0\,.
\ee
For the rational part in $e^{i\phi}$ and $S$, clearly the value at the endpoint of the analytic continuation does not depend on the path, so that we can simply set $e^{i\phi}\to -e^{i\phi}, S\to -S$, and in fact all extra minus signs always cancel out. The HPLs with positive weights also remain unchanged, since they only have branch cuts for their argument between 1 and infinity, and the path (\ref{eq:analytic_cont}) never crosses them for $S>0$. Therefore analytically continuing our expressions amounts to the almost trivial replacement
\be
\log S\to\log S+i \pi\,, \,\text{ or }\,\,\, \sigma\to\sigma+i \pi\,.
\ee
The final step is to pass from the analytically continued collinear to multi-Regge kinematics. As was first noticed in \cite{Dixon:2013eka}, and can be readily verified from (\ref{crossratios}) and (\ref{MRLvars}), in the variables we're currently using MRK corresponds to $T\to 0, S\to 0$ with $r=T/S$ fixed. The ratio $r$ has already appeared in (\ref{wwstar}), and it is in fact via this procedure that the relation with the $w, w^*$ variables of the MRK appearing on the left-hand side of the latter formula can be established.

We thus arrive at the following connection between the two kinematics regions: Starting with the analytically continued collinear limit expansion (around $T=0$), expanding additionally around $S=0$ and replacing $T=r S$ lands us on the multi-Regge limit expansion (around $S=0$), where we are additionally expanding around $r=0$. It is specifically the non-power suppressed $\mathcal{O}(S^0)$ term that has principally been the focus of BFKL analysis, and also of our paper here.

For that reason, we will take the strict $S\to 0$ limit of our expressions (up  to large logarithms), under which we see that, very interestingly, all but the $k=0$ part of all terms (\ref{collexp_genform}) already drops out. Further examining the $k=0$ term for $\mathcal{W}^{(l)}_{2[1,1]}$, $\mathcal{W}^{(l)}_{2[2,1]}$ up to $l=5$, and also $\mathcal{W}^{(4)}_{2[3,1]}+\mathcal{W}^{(4)}_{2[2,2]}/2$, we see that it vanishes in all cases. We deem this as very compelling evidence that 2-particle gluon bound states do not contribute in MRK in general, since this behaviour remain unchanged across loop orders, and also for qualitatively quite different types of integrals.

\section{Multiple Polylogarithms}\label{appx_MPL}

In this appendix we review the definitions and several properties of multiple polylogarithms (MPLs), which will be useful throughout the main text. More detailed expositions may be found for example in \cite{Weinzierl:2010ps,Duhr:2014woa,Panzer:2015ida}.

Multiple polylogarithms may be defined as nested sums\footnote{Note that a different convention with the order of the summation indices reversed, i.e. $i_k>\ldots>i_1$, also exists in the literature.}
\be\label{Li_def}
\text{Li}_{m_1,\ldots,m_k}(x_1,\ldots,x_k)\equiv\sum_{i_1>i_2>\ldots>i_k>0}^\infty \frac{x_1^{i_1}}{i^{m_1}}\cdots\frac{x_k^{i_k}}{i_k^{m_k}}\,,
\ee
in the region
\be\label{Li_region}
|x_1 x_2\ldots x_j|\le 1\quad\forall j=1,\ldots k\,,\quad\text{with }(m_1,x_1)\ne (1,1)\,,
\ee
so that the above series converges. Comparing (\ref{Li_def}) with the definition of $Z$-sums (\ref{Zsum}), it is evident that MPLs are special cases of the latter with the outer summation index set to infinity, $\text{Li}_{m_1,\ldots,m_j}(x_1,\ldots,x_j)=Z(\infty;m_1,\ldots,m_j;x_1,\ldots,x_j)$. Thus similarly to the discussion of $Z$-sums around eq.~(\ref{qshuffle_example}), rearranging the summation ranges of a product of MPLs gives rise to \emph{quasi-shuffle} relations equating the product to a linear combination of MPLs. For example,
\be
\text{Li}_{m_1}(x_1)\text{Li}_{m_2}(x_2)=\text{Li}_{m_1,m_2}(x_1,x_2)+\text{Li}_{m_2,m_1}(x_2,x_1)+\text{Li}_{m_1+m_2}(x_1 x_2)\,.
\ee

The nested sum definition (\ref{Li_def}) is more suited for the evaluation of Mellin-Bernes type integrals by residues, as we did in section \ref{DSOPE}. However there also exists an alternative, global definition of MPLs in terms of (regularised) iterated integrals, as
\be\label{G_def}
G(a_1,\ldots,a_n;z)\equiv\begin{cases}
\frac{1}{n!}\log z&\text{if}\quad a_1=\ldots=a_n=0\\\\
\int_0^z \frac{dt_1}{t_1-a_1}G(a_2,\ldots,a_n;t_1)\,,\quad G(;z)=1\,, &\text{ otherwise.}
\end{cases}
\ee
Unfolding the generic case of the second line for $a_k\ne 0$, we may equivalently write it as
\be\label{Gexp_def}
G(a_1,\ldots,a_n;z)=\int_0^z \frac{dt_1}{t_1-a_1}\int_0^{t_1} \frac{dt_2}{t_2-a_2}\cdots \int_0^{t_{n-1}} \frac{dt_n}{t_n-a_n}\,.
\ee
$\vec a\equiv (a_1,\ldots,a_n)$ is usually denoted as the singularity vector, and its length $n$ as the weight or transcendentality of the MPL.

Although seemingly different, the sum and integral definitions 
are equivalent in the region (\ref{Li_region}), and in particular
\be\label{LitoG}
\text{Li}_{m_1,...,m_k}(x_1,...,x_k)
 = (-1)^k G(\underbrace{0,...,0}_{m_1-1},\frac{1}{x_1},\underbrace{0,...,0}_{m_2-1},\frac{1}{x_1 x_2},...,\underbrace{0...,0}_{m_k-1}{,\frac{1}{x_1...x_k}};1)\,.
\ee
A useful identity which follows immediately from (\ref{Gexp_def}), is that for $a_k\ne0$ $G$-functions are invariant under an arbitrary rescaling $x\in\mathbb{C}^*$ of all the arguments, namely 
\be\label{Grescale}
G(a_1,...,a_k;z) =  G(x a_1, ..., x a_k; x z)\,.
\ee

Finally, the integral definition of MPLs reveals the existence of further \emph{shuffle algebra} relations between products of $G$-functions with the same rightmost argument, such as
\be
\begin{aligned}
G(a;z)G(b;z)&=\int_0^z \frac{dt_1}{t_1-a}\int_0^z \frac{dt_2}{t_2-b}\\
&=\int\int_{0\le t_2\le t_1\le z}\frac{dt_1}{t_1-a} \frac{dt_2}{t_2-b}+\int\int_{0\le t_1\le t_2\le z}\frac{dt_1}{t_1-a} \frac{dt_2}{t_2-b}\\
&=G(a,b;z)+G(b,a;z).
\end{aligned}
\ee
More generally, 
\be
G(\vec a;z)G(\vec b;z)=\sum_{\vec c\in\vec a\sha \vec b}G(\vec c;z)\,
\ee
where the shuffle product $\vec a\sha \vec b$ is defined as the set of all permutations of the elements of $\vec a\cup \vec b$, that preserve the ordering among the elements $a_i\in \vec a$, and among the elements $b_i\in \vec b$.

Interesting special cases of MPLs include the harmonic polylogarithms \cite{Remiddi:1999ew}, for singularity vector entries $a_i\in\{0,\pm 1\}$, and the two-dimensional harmonic polylogarithms \cite{Gehrmann:2000zt}, for $a_i\in\{0,1,-w,1-w\}$, a subset of which we have encountered in this paper\footnote{In the notation of \cite{Gehrmann:2000zt}, 2dHPLs are also defined with an extra minus sign when the sum of all the $a_i$ equal to 1 or $1-z$ is odd.}.

\bibliographystyle{JHEP}
\bibliography{heptagon}

\end{document}